# Ecology of Near-Earth Space

E. H. Nikoghosyan

*NAS RA V. Ambartsumian Byurakan Astrophysical Observatory (BAO)*
*E-mail: elena@bao.sci.am*

## Abstract

The technical achievements of our civilization are accompanied by certain negative consequences affect the near-Earth space. The problem of clogging of near-Earth space by "space debris" as purely theoretical arose essentially as soon as the first artificial satellite in 1957 was launched. Since then, the rate of exploitation of outer space has increased very rapidly. As a result, the problem of clogging of near-Earth space ceased to be only theoretical and transformed into practical.

Presently, anthropogenic factors of the development of near-Earth space are divided into several categories: mechanical, chemical, radioactive and electromagnetic pollution.

## 1. Introduction

Development of our civilization is always inevitably crossed with the questions of preservation of natural and favourable conditions for our life. The mankind aspires forward, mastering all new and new spaces, including, near-Earth. Achievements of scientific and technical progress, including the launch and operation of spacecrafts, entered very strongly in our life. However, the technical achievements of our civilization are accompanied by certain negative consequences, which, first of all, affect the environment around us, including near-Earth space. This contributed to the origin of one of the youngest areas of science - ecology, the science of the relationship of organisms, their communities and the environment. One of the directions of ecological researches is studying of the consequences of our activity in near-Earth space. Unfortunately, it is not always possible to evaluate accurately a real situation, and furthermore, to calculate the consequences. In many cases we can assess current situation only statistically. What can be said for sure is that the environment surrounding us is an interconnected mechanism and that an unreasonable exploitation of which can lead to irreversible consequences.

## 2. Near-Earth space

Near-Earth space (NES) is the global environment surrounding the biosphere of our planet. The zone of its action modern authors define differently, depending on the tasks they solve. Many researchers consider that NES can be prolonged up to the Earth incidence border that makes about 930 000 km. Most often it is the area from the layers of the neutral terrestrial atmosphere (160-200 km) up to the lunar orbit, which is about 384,400 km.

The composition of the NES includes the upper layers of the atmosphere, the ionosphere, and the magnetosphere with radiation belts. It is penetrated by gravitational, geomagnetic, geoelectric and interplanetary magnetic fields, solar wind, streams of charged particles of solar and galactic origin. Comets, asteroids and their fragments, meteor showers, interplanetary dust, etc. fall on it. The interaction of the components of the NES with each other causes complex exchange processes which exert both direct and indirect influences on the biosphere of the Earth, affecting to a certain extent the course of physical, biological, evolutionary processes in animate and inanimate nature.

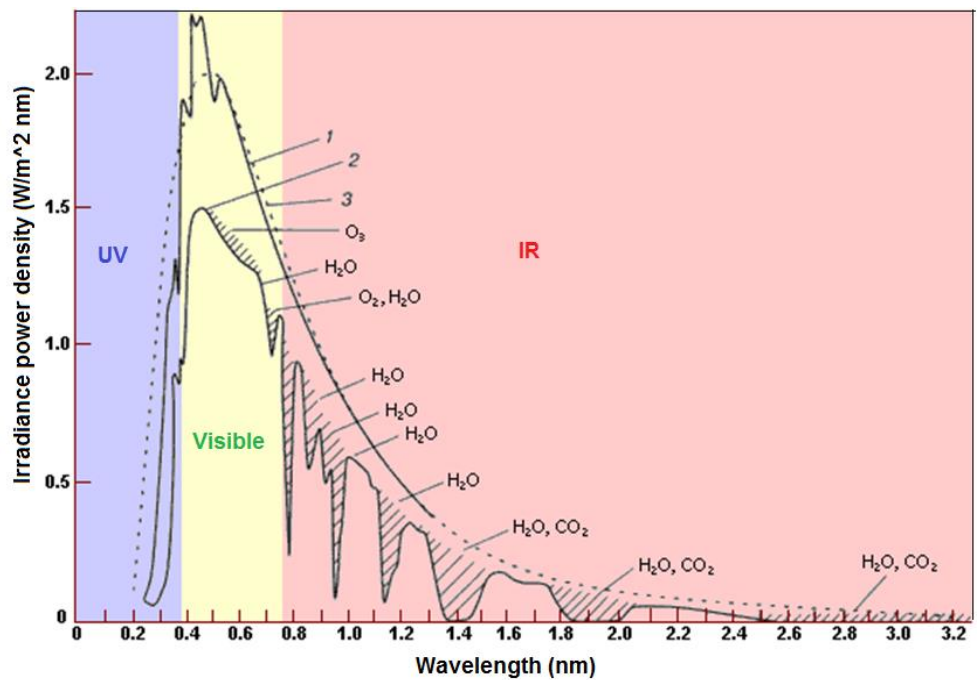

**Figure 1.** Solar spectra in UV, visible and IR ranges: 1 - outside the atmosphere; 2 - on a sea level; dashed line - the radiation of a black body with a temperature of 6000° K. The shaded areas show the absorption by the atmospheric constituents.

Undoubtedly, the main energy supplier in NES is the Sun, under the influence of which the overwhelming number of processes in the Earth-NES system take place. The spectra of the solar radiation is shown on Fig. 1 (Муртазов, 2008). In the visible and infrared wavelengths it is close to the spectrum of a black body with a temperature of 6000° K. The solid curves 1 and 2 in Fig. 1 show the spectra of the solar radiation outside the Earth's atmosphere and at sea level. In the second case, the radiation intensity is reduced due to an absorption in the atmosphere. Due to selective absorption (shaded areas) by the atmospheric constituents ($O_2$, $H_2O$, $CO_2$, etc.), the radiation attenuation is uneven. Solar ultraviolet radiation in 200 - 400 nm range is biologically active, i.e. has a high ability to affect organisms and the biosphere. This range is usually divided into two intervals: 320-400 nm (UV-A) and 200-320 nm (UV-B). Only UV-A radiation reaches the surface of the earth. The harmful to all living matter UV-B radiation is almost completely absorbed by ozone ($O_3$).

The atmosphere around the Earth not only provides the air we need, but also performs a protective function. It has a complex layered structure. The lowest layer (troposphere), which extends to a height of 12-15 km, contains 90% of the mass of the entire atmosphere. The ozone layer, which is currently the subject of active debate, locates at an altitude of 10 to 60 km, with a maximum concentration at an altitude of 20 - 25 km. Let's note that the percentage of ozone is a fraction of a percent. If to distribute the ozon evenly on all atmosphere, then its thickness will be less than 3 mm. But ozone substantially provides absorption of harmful UV. In addition, atomic oxygen and nitrogen are also involved in this process.

Above an altitude of 30 to 1000 km is the ionosphere, named so, because the substance there is in an ionized state. It is due to the ionosphere, such a need for us a distant radio shortwave is carried out. In addition, the ionosphere, and to a greater extent the uppermost layer - the magnetosphere, regularly protect us from penetrating into the lower layers of the atmosphere of high-energy cosmic rays.

In addition to protecting from short-wave radiation harmful to humans, the atmosphere also protects us from the penetration of meteoric bodies and the debris of spacecrafts, which in total is called natural and technogenic space debris. Bursting with great speed into the atmosphere, fragments of space debris create local regions with an increased degree of ionization.

Despite the great distance, there are many undeniable facts that all changes in the atmosphere lead to certain consequences on the surface of the Earth and vice versa. According to modern estimates, the changes in physical parameters (for example, electron concentration), which human activity brings in this area, is still small. They are almost two less, than natural. However, the scale of human activity is growing exponentially. Furthermore, the density in the upper atmosphere is very small. If the molecular density in air in the upper layers of the troposphere is $3 \cdot 10^{19}$ particles/cm$^3$, the density at a height of 100 km is $3 \cdot 10^{13}$ particles/cm$^3$, then at an altitude of 300 km it already drops to $3 \cdot 10^9$ particles/cm$^3$. Because of the high sparseness of matter, any changes in the upper layers of the atmosphere are restored much more slowly than in the lower layers. For example, the increase in the radiation background in the uppermost layers of the atmosphere - the magnetosphere, which was formed as a result of a series of nuclear explosions conducted on the surface of the Earth in the 1960s, lasted more than 10 years.

### 3. Ecology of near-Earth space

In 1957, with the launch of the first artificial satellite was launched near-Earth space exploration. Presently, it is very difficult for us to imagine our life without satellite technology. According to the functional purpose, satellites are classified into the following categories: scientific, geodesic, meteorological, navigational, military and engineering.

The problem of clogging of near-Earth space by "space debris" as purely theoretical arose essentially as soon as the first artificial satellites were launched. At present, as a result of active exploitation, the problem of clogging of near-Earth space ceased to be only theoretical and transformed into practical. Anthropogenic factors of the development of near-Earth space are divided into several categories: mechanical, chemical, radioactive and electromagnetic pollution.

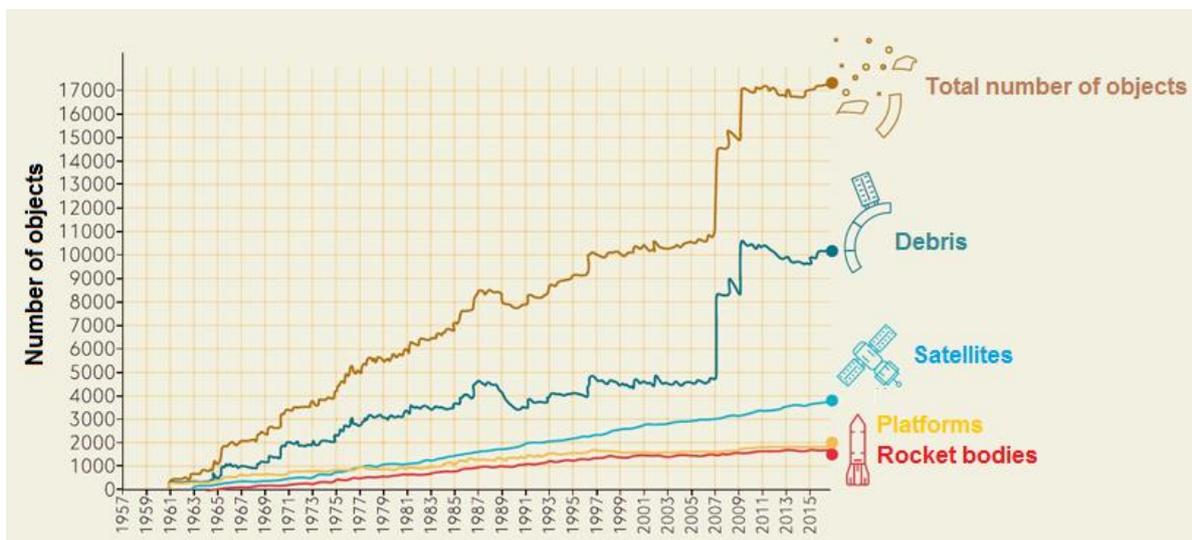

**Figure 2.** History of pollution of near-Earth space by objects of different categories since 1957 (Вениаминов, 2016).

*Mechanical pollution*

According to the data of 2015, over 17,000 space objects of artificial origin are catalogued on near-earth and geostationary orbits. Their total mass is close to 7000 tons. On Fig. 2 the distribution of the total number of registered artificial objects of various categories is shown. The graphs clearly show that the number of fragments over the past years has grown very rapidly. The part, suppressing in a percentage ratio is the share of fragments or scraps. Of the total number of objects, only about 6% are "active", i.e. functional satellites. The remaining 94% can be attributed to space debris (see Fig. 3), which does not perform any useful work. It should be noted that the data represent only the registered objects. Their real number is undoubtedly greater.

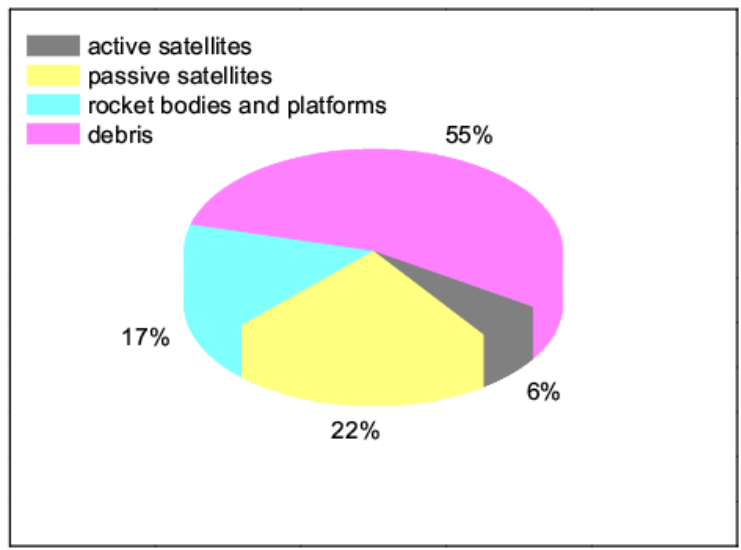

**Figure 3.** The composition of artificial space bodies.

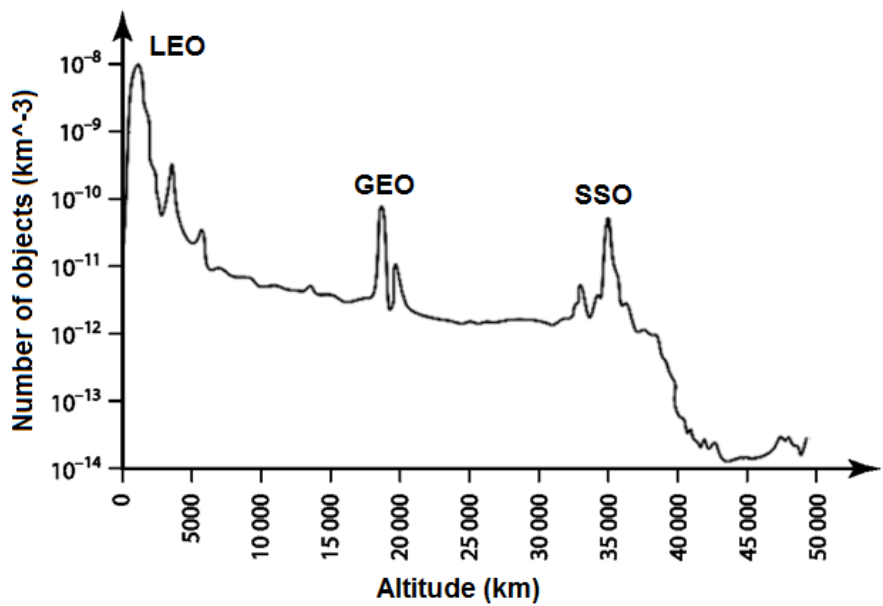

**Figure 4.** The height distribution of catalogued artificial space objects (data from Kaman Sciences Corporation).

Artificial, technogenic bodies are not equally distributed in the near-Earth space. Most clogged those of Earth's orbit, which are most often used for spacecrafts. These are Low Earth orbits (LEO), Geostationary orbits (GEO) and Sun-Synchronous orbits (SSO).

Presently, as a result of human activity in near-Earth space, the flow of fragments of artificial origin in LEOs is comparable to the flow of bodies of natural origin (meteoritic matter), which is reflected in Fig. 5 (Новиков, 2006). It should be noted that the figure shows the data only for objects with a diameter greater than 10 cm.

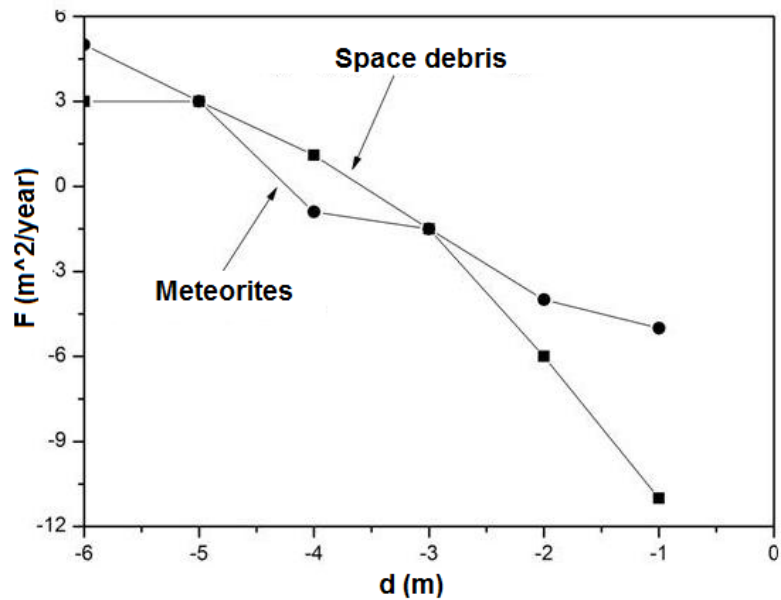

**Figure 5.** The ratio of flows of technogenic objects and meteoroid matter of various sizes in the region of low near-Earth orbits.

In addition to the "large" particles in the universe, there are many smaller particles that emerge during explosions on satellites or as a result of collision with the meteorites. According to various estimates, the number of fragments with 1 -10 cm diameter reaches 200 - 250 thousands, and with 0.1 - 1 cm range from 80 to 100 millions. Viewing and calculating the objects with these dimensions are very difficult. In addition, there is a very large number (about 1016) of particles with a diameter of 1 to 10 microns, which are fragments of solid propellant, paint, etc. Although their lifetime is relatively small and makes up to 1 to 10 days, their number (mainly due to clashes) is rapidly growing and at present the density of these small particles is about 4 g/cm$^3$.

### *Chemical pollution*
Chemical pollution is mainly caused by the release of rocket fuel into the atmosphere. This has a much greater effect on the lower layers of the atmosphere than on near-Earth space. In general, this occurs at an altitude of 10-40 km, in the region of the location of the ozone layer. However, in defense of rocket technology, it should be noted that this is an insignificant part, in relation to the anthropogenic impacts of industry. In addition, the release of rocket fuel also affects the ionosphere, forming, at an altitude of 400-500 km, so-called ionospheric holes, i.e. regions with a lower electron concentration, which affects the propagation of radio emission and creates interference in radio communication lines.

*Electromagnetic pollution*

Electromagnetic pollution is called the technogenic radiation that created by satellites and their transmitters. Penetrating in the ionosphere and magnetosphere the electromagnetic pollution also causes environment ionization degree changes, which, in turn, can affect the quality of radio communication.

*Radioactive pollution*

Radioactive pollution occurs due to the fact that in certain satellites the radioactive substances is used. These substances can penetrate in the atmosphere and even reach the Earth's surface.

## 4. Prediction of the ecological situation in the near-Earth space

Thus, the achievement of cosmic space, the exploitation of space equipment, has dramatically widened the range of environmental issues. We will never abandon our technical achievements. On the contrary, there are all reasons to assume that the rate of exploitation of outer space will be increasing. That is why we should be able to make accurate estimates of the inevitable consequences and take timely action to prevent them, or at least mitigate them.

Unfortunately, at present we can only control the near-Earth space. The monitoring includes observations and constant control of the general condition, the degree of natural and anthropogenic pollution. It is also necessary to develop methods for assessing the physical state of near-Earth space as part of the natural environment, both for a given period of time and in the future. To assess the state of near-Earth space in the near future computational models are created: Russian SDPA (Space Debris Prediction and Analysis), European MASTER (Meteoroid and Space debris Terrestrial Environment Reference model), American ORDEM (Orbital Debris Engineering Model), etc. Fig. 6 shows the results of the forecast for the next 50 years, made with SDPA model. This figure shows the relative change in the total number of technogenic space objects larger than 1 cm in the region of LEOs (below 2000 km). In the simulations 5 different scenarios of near space operation were used:

1. The intensity of clogging of near-Earth space will remain at the level of the previous 10 years;
2. Scenario 1 plus the exclusion of the accompanying fragments of the launching rockets;
3. Scenario 1 plus the exclusion of explosions of artificial space objects;
4. Decreasing by 2 times the number of the launches of spacecraft;
5. Simultaneous application of the scenarios 2 - 4.

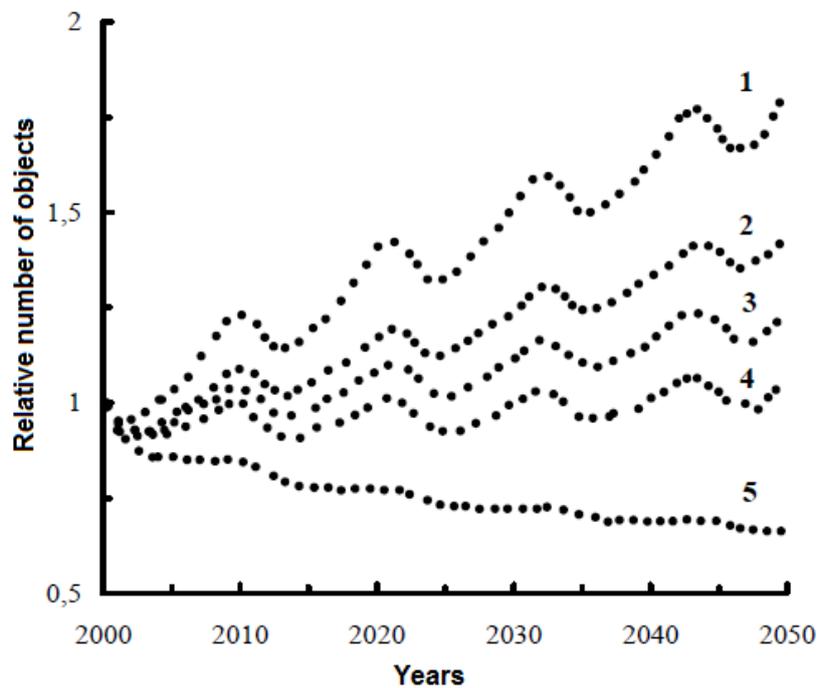

**Figure 6.** A forecast of the total number of technogenic space objects larger than 1 cm in LEOs (Муртазов, 2008).

On graphics it is visible that at scenario 1-4 is predicted the growth of the number of man-made objects. Periodic fluctuations in the number of objects on the graph are associated with the 11-year cycle of solar activity, which causes modulation of the average density of the upper atmosphere of the earth and, accordingly, increased braking of space objects during the period of maximum solar activity. First option - the most unfavourable. In accordance with it, in 2050 the number of man-made objects will increase by 1.8 times. The most effective method is to exclude explosions of spacecraft. Under scenario 5, the probability of reducing the level of pollution of near-Earth space in 2050 will be 25-30%.

Figure 7 shows the forecasts of the number of collisions per year between catalogued objects for the following scenarios:

1. Preservation on the same level the intensity of launches;
2. Preservation on the same level the intensity of launches during 20 years and their further termination;
3. Termination of further launches.

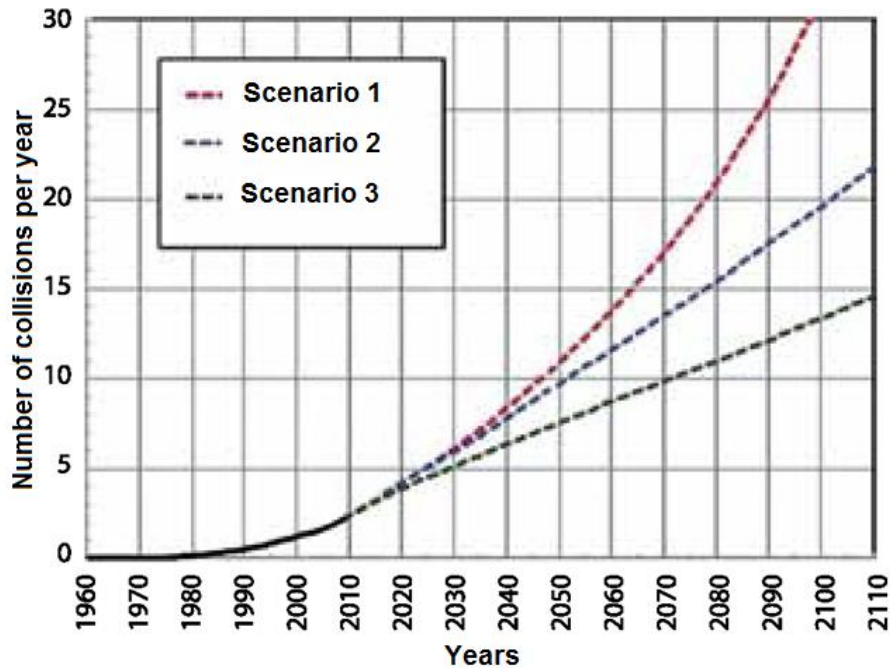

**Figure 7.** Forecasts of the number of collisions per year between catalogued objects.

In all cases, the forecasts are not comforting, especially taking into account the fact that there is no reason to expect a reduction in the number of launches of new spacecrafts. There is a need to develop ways to clean fragments already flying in space. Similar developments are already under way. However, their implementation requires very large investments and joint, international efforts.

We want, nevertheless, to finish on an optimistic note. Mankind is not the first time to face global problems and the way out has always been.

**References:**
Bairds, J. C. 1989, The Inner Limits of Outer Space
Encountering Life in The Universe, eds. Impey, C.; Spitz, A. H.; Stoeger, W., 2012;
Morton, T. 2016, Dark Ecology: For a Logic of Future Coexistence (The Wellek Library Lectures)
Вениаминов, С. 2016, журнал ВКС, N 1, стр. 86
Новиков, Л. С. 2006, Основы экологии околоземного космического пространства, Москва

Муртазов, А. К. 2008, Физические основы космического пространства, Рязанский Государственный Университет